\documentclass[a4paper, 10pt]{article}
\usepackage[utf8]{inputenc}
\usepackage{color}
\usepackage{hyperref} 
 \usepackage{latexsym}
  \usepackage{amssymb}
  \usepackage{graphicx}
   \DeclareGraphicsExtensions{.jpg .png .gif .pdf .pdf}
\usepackage[T1]{fontenc}
 \usepackage[english]{babel}
  \usepackage{}
  \usepackage{amstext}
  \usepackage{amsmath}
  \usepackage{times} 
  \usepackage{amsfonts}
 \date{}

\title{Effect of nonlinear dissipation on the basin boundaries of a driven two-well or catastrophic single-well Modified Rayleigh-Duffing Oscillator }
\author{C. H. Miwadinou\footnote{clement.miwadinou@imsp-uac.org, hodevewan@yahoo.fr},   
A. V. Monwanou\footnote{movins2008@yahoo.fr} and 
J. B. Chabi Orou\footnote{Author to whom correspondence should be addressed: jchabi@yahoo.fr}}

\begin{document}

\maketitle

\begin{abstract}
This paper considers effect of nonlinear dissipation on the basin boundaries of a diven two-well Modified Rayleigh-Duffing Oscillator
where pure and unpure quadratic and cubic nonlinearities are considered. By analyzing the potential
an analytic expression is found for the homoclinic orbit. The Melnikov criterion is used to examine
a global homoclinic bifurcation and transition to chaos in the case the of our oscillator. It is found the effect of unpure quadratic parameter and 
amplitude of parametric excitation on the critical Melnikov amplitude $\mu_{cr}$. Finally, we examine carefully the phase space of initial conditions
in order to analyze the effect of the nonlinear damping, and particular how the basin boundaries become fractalized.
\end{abstract}

{\bf keywords}: Catastrophic single-well potential, two-well potential,\\
 modified Rayleigh-Duffing oscillator, melnikov criterion, bifurcation, 
chaotic behavior, basin boundaries.

\section{Introduction}
The Rayleigh oscillator is one canonical example of self-excited systems. However, generalizations of such
systems, such as the Rayleigh–Duffing oscillator, have not received much attention. The presence of a pure and unpure quadratic and cubic terms makes
the Rayleigh–Duffing oscillator a more complex and interesting case to analyze. This oscillator is used to modelize the following phenomenon: 
a El Ni$\tilde{n}$o  Southern Oscillation $(ENSO)$ coupled tropical ocean-atmosphere weather phenomenon  in which the state
variables are temperature and depth of a region of the ocean called the thermocline (where the annual seasonal cycle is the parametric excitation and the 
model exhibits a Hopf bifurcation in the absence of parametric excitation),  a $MEMS$ device  consisting 
of a $30\mu m$ diameter silicon disk which can be made to vibrate by heating it with a laser beam resulting in a Hopf bifurcation
 (where the parametric excitation is provided by making the laser beam intensity vary periodically in time) etc. \cite{16} and \cite{18}. 

The behavior of Rayleigh-Duffing oscillator with periodic forcing and/or parametric excitations  has been investigated extensively by many researchers. 
For instance,  in their work, the siewe siewe and their collaborators \cite{20} have studied the nonlinear response and suppression of chaos by weak harmonic
 perturbation inside a triple well $6-$Rayleigh oscillator combined to parametric excitations. Three years ago, Siewe Siewe and al.\cite{2} investigated 
the effect of the nonlinear dissipation on the boundaries of a driven two-well Rayleigh-Duffing oscillator and in other paper \cite{3}, the same authors 
focussed their analyze on the occurrence of chaos in a parametrically driven extended
Rayleigh oscillator with three-well potential.  However, in many situations, the nonlinear dynamics dominates the behavior of physical systems giving rise to
multi-stable potentials or catastrophic monostable potentials. The authors in Refs. \cite{2,3} and \cite{20} showed with a rigorous theoretical consideration 
that the resonant parametric perturbation
can remove chaos in low dimensional systems. They confirmed this prediction with numerical simulations.  It is interesting to note that there is a situation
analyzed in \cite{21}, where Melnikov analysis is applied to a nonlinear oscillator which can behave as a one-well oscillator, a two-well oscillator
or three-well oscillator by simply modifying one of its parameters, which acts as a symmetry-breaking mechanism. Therefore, the chaotic behavior using the 
parametric perturbation in the modified Rayleigh–Duffing oscillator with a two-well potential still needs to be investigated further. Another good example
is constituted by the generalized perturbed pendulum \cite{22}.
Our aim is to  make a contribution in the study of the transition to chaos in the modified Rayleigh–Duffing oscillator by using the Melnikov theory, and then
 see how the fractal basin boundaries arise and are modified as the damping coefficient is varied. The last part of this work consists of a numerical 
investigation of the strange attractor at parameter values which are close to the analytically predicted bifurcation curves. In particular, the case of the
 two-well potential is considered.

The paper is organized as follows. In the next section, after describing the model, analyzing of the model, and some comparison with the simple 
Rayleigh-Duffing oscillator, the conditions for the existence of chaos are thoroughly analyzed. A convenient demonstration of the accuracy of the 
method is obtained from the fractal basin boundaries, and this is discussed in Section 4. We conclude in the last section. In the
appendixe $ A$ we show the Melnikov integration procedures in details.

\section{Desciption and analysis of the model}
In this paper, we examine the dynamical transitions in parametric and periodically forced self-oscillating systems
containing the cubic terms in the restoring force and the pure and hybrid quadratic and cubic
in nonlinear damping function as follows:
\begin{eqnarray}
 &&\ddot{x} + \epsilon\mu (1-\dot{x}^2)\dot{x}+\epsilon\beta\dot{x}^2+\epsilon k_1\dot{x}x+\epsilon k_2\dot{x}^2 x +(\gamma+\alpha\cos\Omega t)x\cr
&&+\lambda x^3=F\cos\Omega t, \label{eq.1}
\end{eqnarray}
where $\epsilon, \mu, \beta, k_1, k_2, \gamma, \lambda, F$ and $\Omega$ are parametrs. Physically, $\mu, k_2, \beta$ and $k_1$ represent respectively pure, unpure cubic and pure, unpure quadratic nonlinear damping
coefficient terms, $ \alpha$ and $F $ are the amplitudes of the parametric and external periodic forcing, and
$\sqrt{\gamma}$ and $\Omega$ are respectively natural and external forcing frequency. Moreover $ \lambda$ characterize the intensity of the nonlinearity and $\epsilon$ is the nonlinear damping parametr
control. The nonlinear damping term corresponds to the Modified Rayleigh
 oscillator, while the nonlinear restoring
force corresponds to the Duffing oscillator, hence its name Modified Rayleigh-Duffing oscillator.

In this section, we derive the fixed points and the phase portrait corresponding to the system Eq. (\ref{eq.1}) when it is
unperturbed. If we let $ \epsilon=\alpha=F=0$, Eq. (\ref{eq.1}) is considered as an unperturbed system and can be rewritten as
\begin{eqnarray}
 \dot{x}=y,\quad  \dot{y}=-\gamma x-\lambda x^3 , \label{eq.2} 
\end{eqnarray}
which corresponds to an integrable Hamiltonian system with the potential function
given by 
\begin{eqnarray}
 V(x) = \frac{1}{2}\gamma x^2+\frac{1}{4}\lambda x^4, \label{eq.3}
\end{eqnarray}
whose associated Hamiltonian function is 
\begin{eqnarray}
 H(x,y)=\frac{1}{2}y^2+ \frac{1}{2}\gamma x^2+\frac{1}{4}\lambda x^4.\label{eq.4}
\end{eqnarray}
From Eqs. (\ref{eq.2})  and (\ref{eq.4}), we can compute the fixed points and analyze their stabilities.

$\bullet$ If $\gamma>0, \lambda>0$ or $\gamma<0, \lambda<0$, the system have one fixed point $(0,0)$ which is a center. 

\begin{figure}[htbp]
\begin{center}
 \includegraphics[width=12cm,  height=6cm]{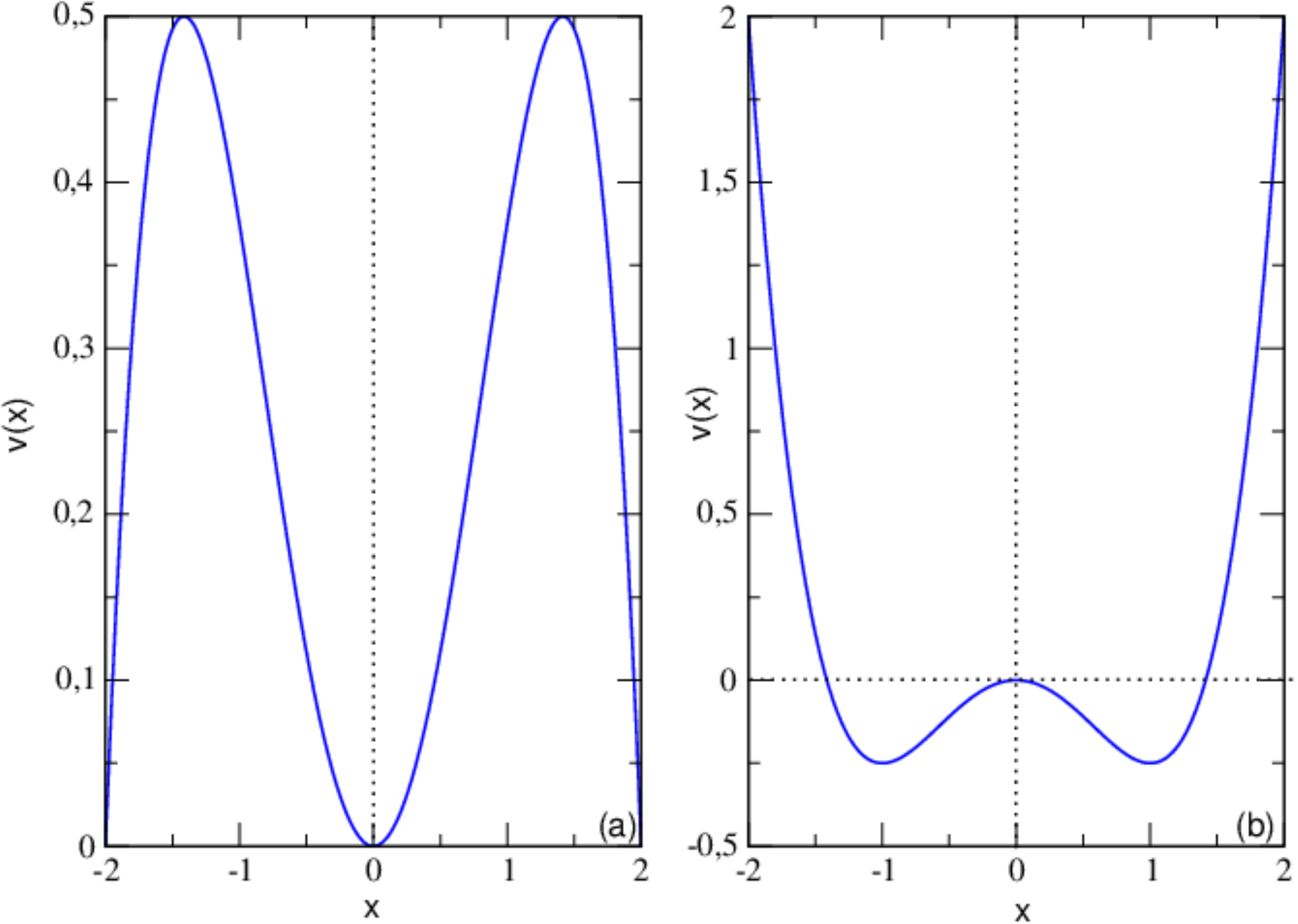}
\end{center}
\caption{$(a)$ The catastrophic single well potential of the unperturbed system (\ref{eq.4}); 
$b$ The two-well potential function of the unperturbed system (\ref{eq.4}).}
\label{fig:1}
\end{figure}

$\bullet$ For $\gamma>0, \lambda<0$, there are three fixed points: two saddles connected
by two heteroclinic orbits and one center. The potential defined by Eq. (\ref{eq.3}) has two-well (see Fig. \ref{fig:1}$(b)$).

$\bullet$ For $\gamma<0, \lambda>0$, there are three fixed points: two saddles connected
by two heteroclinic orbits and one center. The potential defined by Eq. (\ref{eq.3}) has catastrophic sigle-well (see Fig. \ref{fig:1}$(a)$ ).

Therefore, depending on the values of the external excitation, the system can escape over the potential
barrier and dramatically suffers an unbounded motion. Fig.  represents the
corresponding phase portraits between the unforced Rayleigh–Duffing oscillator and the unforced Modified Rayleigh–Duffing oscillator,
respectively with single-well (left) and two-well (right) conditions.

\begin{figure}[htbp]
\begin{center}
 \includegraphics[width=12cm,  height=6cm]{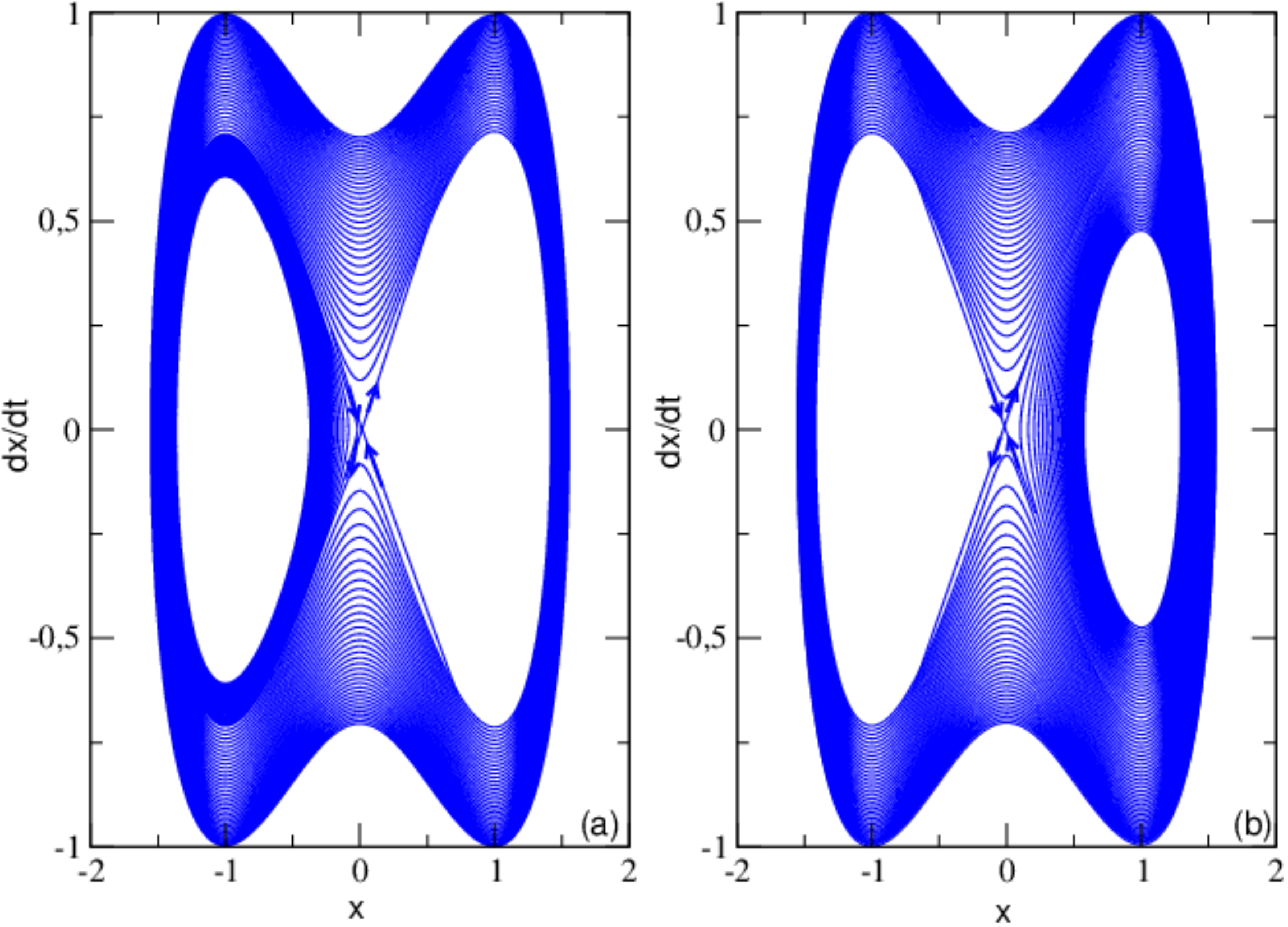}
\end{center}
\caption{ $(a)$ Phase portrait of the unforced Rayleigh-Duffing oscillator with $\gamma=-1, \lambda=1,\alpha=0,
 \beta=0, \mu=0.2, \epsilon=0.1, k_1=0, k_2=0$; $(b)$ Phase portrait of the unforced Modified Rayleigh-Duffing 
oscillator with $\gamma=-1, \lambda=1,\alpha=0.5, \beta=0.5, \mu=0.2, \epsilon=0.1, k_1=0.5, k_2=0.5$}
\label{fig:2}
\end{figure}

\newpage

\section{Taming chaotic behavior in the Modified Rayleigh-Duffing oscillator}
 In this section, we discuss the chaotic behavior of the system
\begin{eqnarray}
 &&\ddot{x} + \epsilon\mu (1-\dot{x}^2)\dot{x}+\epsilon\beta\dot{x}^2+\epsilon k_1\dot{x}x+\epsilon k_2\dot{x}^2 x +(\gamma+\alpha\cos\Omega t)x\cr
&&+\lambda x^3=F\cos\Omega t, \label{eq.5}
\end{eqnarray}
where $\mu, k_1, \beta, k_2, \gamma, \alpha, \lambda, \Omega$ and $F$ are assumed to be small parameters. Hence, our dynamical system may be written as
\begin{eqnarray}
 \dot{x}=y,\quad  \dot{y}=-\gamma x-\alpha x\cos\Omega t-\lambda x^3-\epsilon\mu (1-\dot{x}^2)\dot{x}-\epsilon\beta\dot{x}^2-\cr
\epsilon k_1\dot{x}x-\epsilon k_2\dot{x}^2 x +F\cos\Omega t, \label{eq.6} 
\end{eqnarray}
where $F=\epsilon F$

When the pertubations are added, the homoclinic orbit might be broken transversely. And then, by the smale-Birkoff Theorem$[19]$, horseshose type chaotic
dynamics may appear. It is well known, that the predictions for the appearance of chaos are limited and only valid for orbits starting at points 
sufficiently close to the separatrix. On the other hand it constitutes a first order perturbation method. Although the chaos does not manifest itself
in the form of permanent chaos, and some sorts of transient chaos may showup. Hower, it manifest itself in terms of fractal basin boundaries, as it was 
shown by[20]. We start our analysis form the unperturbed Hamiltonian Eq. (\ref{eq.4}). The potential $V(x)$ Eq. (\ref{eq.3}) has the local 
peak (Fig. \ref{fig:1} $(b)$) or local antipeak (see Fig. \ref{fig:1} $(a)$) at the saddle point $x=0$. Esistence of this point with a horizontal tangent
makes possible homoclinic or heteroclinic bifurcations  to take place.\\
At the saddle point $x=0$, for an unperturbed system ( Fig. \ref{fig:1}), the system velocity reaches zero in velocity $y=0$ (for infinite time $t=\pm\infty$)
so the total energy has only its potential part. In this paper, the homoclinic case is be study.

Transforming Eqs. (\ref{eq.3}, \ref{eq.4}), for a closen nodal energy $(H=0)$ and for $\gamma<0, \lambda>0$ we  get the following expression for velocity:
\begin{eqnarray}
 y=\frac{dx}{dt}=\sqrt{2(-\frac{\gamma}{2}x^2-\frac{\lambda}{4}x^4)}. \label{eq.7}
\end{eqnarray}
Now one can perform integration over $x$:
\begin{eqnarray}
 t-t_0=\pm\int\frac{dx}{x\sqrt{-\gamma-\frac{\lambda}{2}x^2}},\label{eq.8} 
\end{eqnarray}
where $t_0$ represents an integration constant. Finally, we get so called homoclinic orbits ( Fig. \ref{fig:2}):
\begin{eqnarray}
&& x_h=\pm\sqrt{\frac{-2\gamma}{\lambda}}sech{(\sqrt{-\gamma}(t-t_0))}\cr
&& y_h=\pm\sqrt{\frac{2}{\lambda}}\gamma sech{(\sqrt{-\gamma}(t-t_0))}\tanh{(\sqrt{-\gamma}(t-t_0))}, \label{eq.9} 
\end{eqnarray}
where $`+`$ and $`-`$ signs are related to left $-$ and righ $+$ sign orbits, respectively ( Fig. \ref{fig:2}). Note, the central
saddle point $x_0=0$ is reached in time $t$ corresponding to $+\infty$ and $-\infty$ respectively.\\
We apply the Melnikov method to our system in order to find the necessary criteria for the existence of homoclinic bifurcations and chaos. 
The Melnikov integral is defined as
\begin{eqnarray}
 M(t_0)=\int_{-\infty}^{+\infty}f(x_h,y_h)\wedge g(x_h,y_h)dt, \label{eq.10} 
\end{eqnarray}
where the corresponding differential form $f$ means the gradient of unperturbed hamiltonian while $g$ is a perturbation from Eq. (\ref{eq.6}) after to 
put $\alpha=\epsilon\alpha$ and $F=\epsilon F$.
Eq. (\ref{eq.10}) can be rewritten as follows:
\begin{eqnarray}
 M^\pm(t_0)&=&-\mu\int y_h^2 dt+\mu\int y_h^4dt-k_1\int x_h y_h^2 dt-\beta\int y_h^3 dt-\cr
&&k_2\int x_hy_h^3 dt-\alpha\int x_hy_h\cos\Omega(t+t_0) dt+\cr
&&F\int y_h\cos\Omega (t+t_0) dt, \label{eq.11} 
\end{eqnarray}

where $t_0$ is the cross-section time of the Poincare map and $t_0$ can be interpreted as the initial time of the forcing term.
After substituting the equations of the homoclinic orbits $x_h$ and $y_h$ given in Eq. (\ref{eq.9}) into Eq. (\ref{eq.11}) and evaluating the
corresponding integral, we obtain the Melnikov function given by
\begin{eqnarray}
 M^\pm(t_0)=-\mu I_0+\mu I_1-k_1 I_2-\beta I_3-k_2 I_4-\alpha\sin{\Omega t_0} I_5+F\sin{\Omega t_0}I_6, \label{eq.12}
\end{eqnarray}
where
\begin{eqnarray}
 I_0&=&\frac{2\gamma^2}{\lambda}\int_{-\infty}^{+\infty}sech^2{(\sqrt{-\gamma}t)}\tanh^2{(\sqrt{-\gamma}t)}dt,\nonumber\\
I_1&=& \frac{4\gamma^4}{\lambda^2}\int_{-\infty}^{+\infty}sech^4{(\sqrt{-\gamma}t)}\tanh^4{(\sqrt{-\gamma}t)}dt,\nonumber\\
I_2&=&\pm\frac{2\gamma^2}{\lambda}\sqrt{\frac{-2\gamma}{\lambda}}\int_{-\infty}^{+\infty}sech^3{(\sqrt{-\gamma}t)}\tanh^2{(\sqrt{-\gamma}t)}dt,\nonumber\\
I_3&=&\pm\frac{2\gamma^3}{\lambda}\sqrt{\frac{2}{\lambda}}\int_{-\infty}^{+\infty}sech^3{(\sqrt{-\gamma}t)}\tanh^3{(\sqrt{-\gamma}t)}dt,\nonumber\\
I_4&=&\frac{4\gamma^3}{\lambda^2}\sqrt{-\gamma}\int_{-\infty}^{+\infty}sech^4{(\sqrt{-\gamma}t)}\tanh^3{(\sqrt{-\gamma}t)}dt,\nonumber\\
I_5&=&\frac{2\gamma}{\lambda}\sqrt{-\gamma}\int_{-\infty}^{+\infty}sech^2{(\sqrt{-\gamma}t)}\tanh{(\sqrt{-\gamma}t)}\sin{\Omega t_0}dt,\nonumber\\
I_6&=&\pm\gamma\sqrt{\frac{2}{\lambda}}\int_{-\infty}^{+\infty}sech{(\sqrt{-\gamma}t)}\tanh{(\sqrt{-\gamma}t)}\sin{\Omega t_0}dt.\label{eq.13}
\end{eqnarray}
After evaluation of these elementary integrals (see Appendix ), the Melnikov function is computed. 
\begin{eqnarray}
&&M^\pm(t_0)=\frac{4\mu\gamma\sqrt{-\gamma}}{3\lambda}-\frac{16\mu\gamma^3\sqrt{-\gamma}}{35\lambda^2} \mp\frac{\pi k_1\gamma^2}{8}(\frac{2}{\lambda})^{\frac{3}{2}}\mp\cr
&&\frac{23}{70}\pi\Omega F\sqrt{\frac{2}{\lambda}}sech(\frac{\pi\Omega}{2\sqrt{-\gamma}})\sin{\Omega t_0}-\frac{\pi\alpha\Omega^2}{2\lambda \sqrt{-\gamma}}cosech(\frac{\pi\Omega}{2\sqrt{-\gamma}})\sin{\Omega t_0}.\label{eq.14} \end{eqnarray}
It is known, that the intersections of the homoclinic orbits are the necessary conditions for the existence of chaos. The Melnikov function 
theory measures the distance between the perturbed stable and unstable manifolds in the Poincaré section. If $M^\pm(t0)$ has a simple zero,
then a homoclinic bifurcation occurs, signifying the possibility of chaotic behavior. This means that only necessary conditions for 
the appearance of strange attractors are obtained from the Poincaré–Melnikov–Arnold analysis, and therefore one has always the chance 
of finding the sufficient conditions for the elimination of even transient chaos. Then the
necessary condition for which the invariant manifolds intersect themselves is given by
\begin{eqnarray}
 &&\mu_{cr}=\frac{\lambda^2}{4\gamma\sqrt{-\gamma}(\frac{1}{3}-\frac{4\gamma^2}{35})}\times\cr
&&\left[ \frac{\pi\alpha\Omega^2}{2\lambda\sqrt{-\gamma}}cosech(\frac{\pi\Omega}{2\sqrt{-\gamma}})
\pm\frac{23}{70}\pi\Omega F\sqrt{\frac{2}{\lambda}}sech(\frac{\pi\Omega}{2\sqrt{-\gamma}})
\pm\frac{\pi k_1\gamma^2}{8}(\frac{2}{\lambda})^{\frac{3}{2}} \right]
\end{eqnarray}
Above this value $\mu\ge\mu_{cr}$ the system transit
through a global homoclinic bifurcation which is a necessary condition for ap-
pearance of chaotic vibrations.

\begin{figure}[htbp]
\begin{center}
 \includegraphics[width=12cm,  height=6cm]{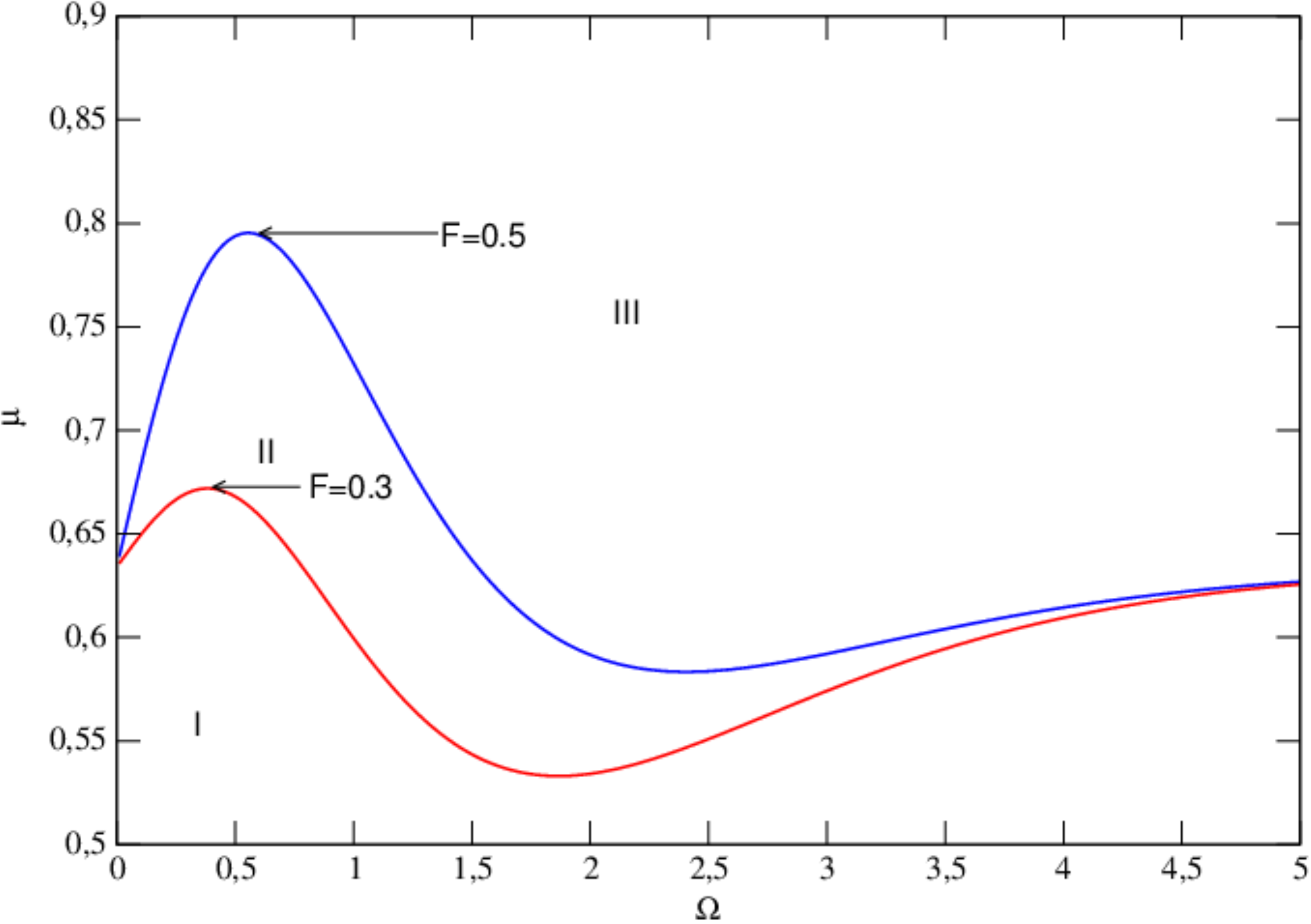}
\end{center}
\caption{ Critical amplitude $\mu$ versus frequency for two different external amplitude parameters values.}
\label{fig:3}
\end{figure}

\begin{figure}[htbp]
\begin{center}
 \includegraphics[width=12cm,  height=10cm]{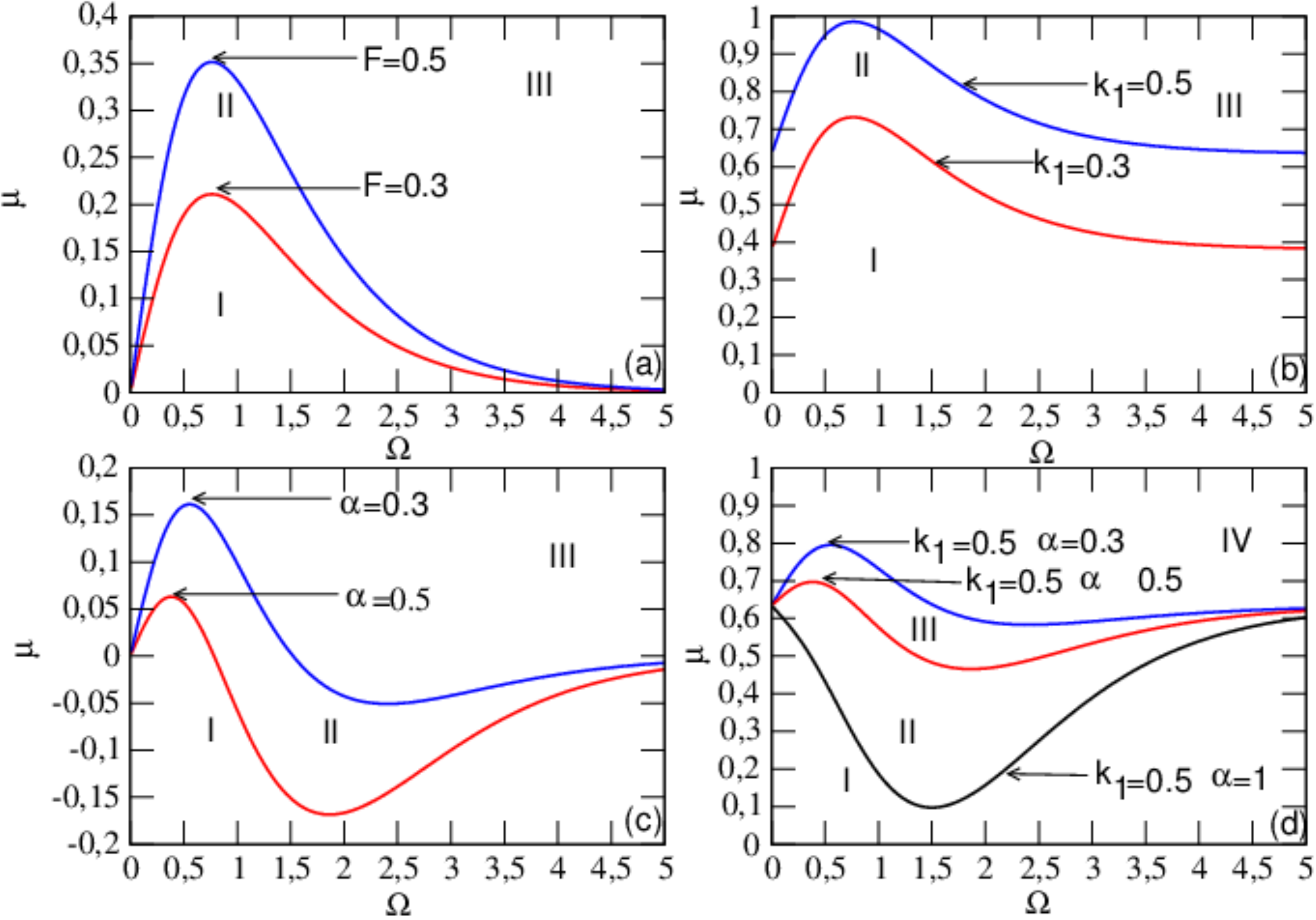}
\end{center}
\caption{ Effect of different amplitude parameters values  on critical amplitude $\mu$ versus frequency showing chaotic regions.}
\label{fig:4}
\end{figure}

This implies that if the perturbation is sufficiently small, the reduced Eq. (\ref{eq.6})  has transverse homoclinic orbits resulting in
possible chaotic dynamics. We study the chaotic threshold as a function of only the frequency parameter $\Omega$. A typical
plot of $\mu$ against $\Omega$ is shown in  Fig. \ref{fig:3} , in which the critical homoclinic bifurcation curves are plotted
 versus the frequency parameter $\Omega$. The threshold of chaotic motion increases with the increasing of the external amplitude 
$\mu$ ( Fig. \ref{fig:3}). The region below the homoclinic bifurcation curve corresponding to $F = 0.3$ (region $(I)$ of Fig. \ref{fig:3}) represents the periodic
orbits. When $\mu$ crosses its first critical value, a homoclinic bifurcation takes place, so that a hyperbolic Cantor set appears in a neighborhood of the saddles
 (regions $(II)$ and $(III)$ of Fig. \ref{fig:3} ). The dynamics should therefore be chaotic only for large values of the damping. At the same time, when $F = 0.5 $
(regions $(I)$ and $(II)$ of Fig. \ref{fig:3} ) it represents the periodic
orbits, while the dynamics should therefore be chaotic in the region $(III)$.
 Fig. \ref{fig:4} represents the effect of different amplitude parameters values  on critical amplitude 
$\mu$ versus frequency showing chaotic regions. When $\alpha=0$ and $k_1=0$ the necessary condition for which the invariant manifolds intersect
themselves corresponding exactly to the condition which is obtained by Siewe Siewe and al. for Rayleigh-Duffing oscillator \cite{2} 
(see Fig. \ref{fig:4} $(a)$). Figs. \ref{fig:4} $(b), (c)$ and $(d)$ show respectively hybrid quadratic nonlinearity, amplitude of excitation parameter
and these two parameters simultanious  effect on critical amplitude $\mu_{cr}$ for $F=0.5$. In these case, we noticed that the parameters $k_1, \alpha$ 
and $F$ are several effect on melnikov critical amplitude which show when chaotic behavior appear in  the modified Rayleigh-Duffing. Critical amplitude 
increases with the increasing of the parameter $k_1$ (see Figs. \ref{fig:4} $(b)$) but with $\alpha$ the critical amplitude are two extrema which show
the effect of parametric excitation.

\newpage

 \section{Bifurcation analysis, phase portraits and fractal basins}
\begin{figure}[htbp]
\begin{center}
 \includegraphics[width=12cm,  height=10cm]{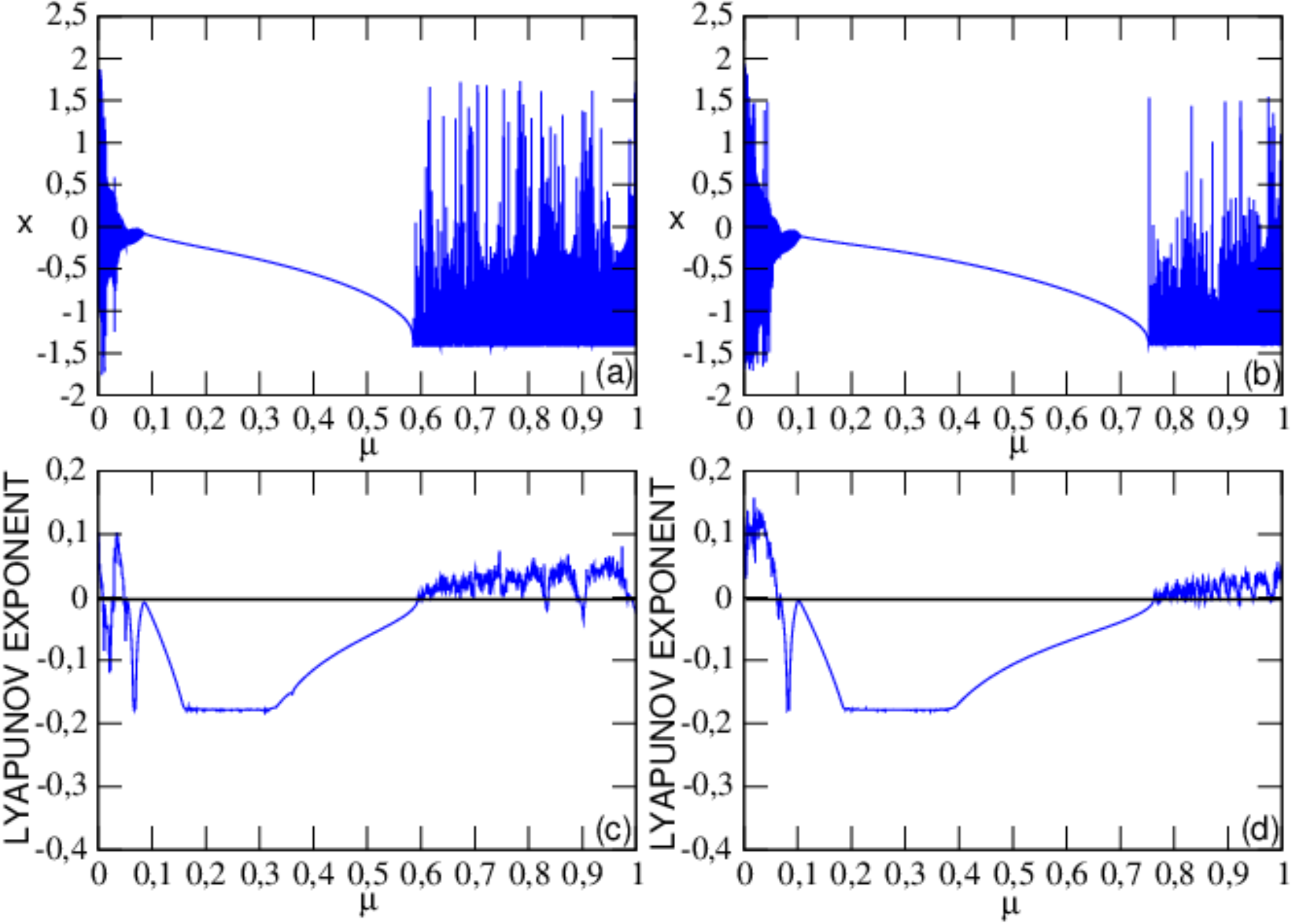}
\end{center}
\caption{ Bifurcation diagram and corresponding Maximal Lyapunov exponent of Modified Rayleigh-Duffing 
oscillator equation versus  $\mu$ with parameters of Fig. \ref{fig:3};  $(a,c) F=0.5 $  , $(b,d) F=0.6$.}
\label{fig:5}
\end{figure}

\begin{figure}[htbp]
\begin{center}
 \includegraphics[width=12cm,  height=10cm]{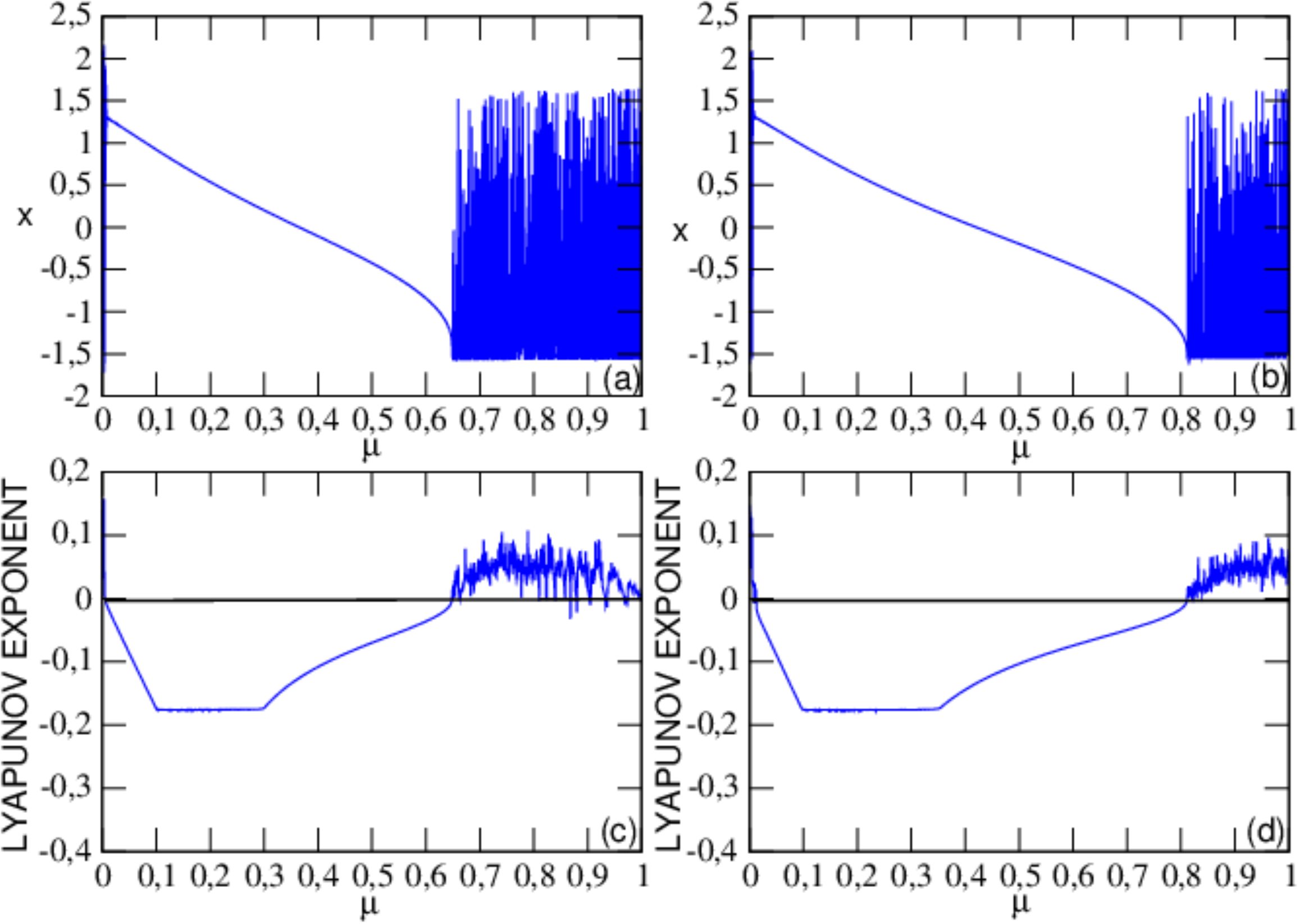}
\end{center}
\caption{ Bifurcation diagram and corresponding Maximal Lyapunov exponent of Modified Rayleigh-Duffing oscillator
 equation versus  $\mu$ when the modified parameters equals to $0$ with parameters of Fig. \ref{fig:4}$(a)$; $(a,c) F=0.5 $, $(b,d) F=0.6$.}
\label{fig:6}
\end{figure}

\begin{figure}[htbp]
\begin{center}
 \includegraphics[width=12cm,  height=10cm]{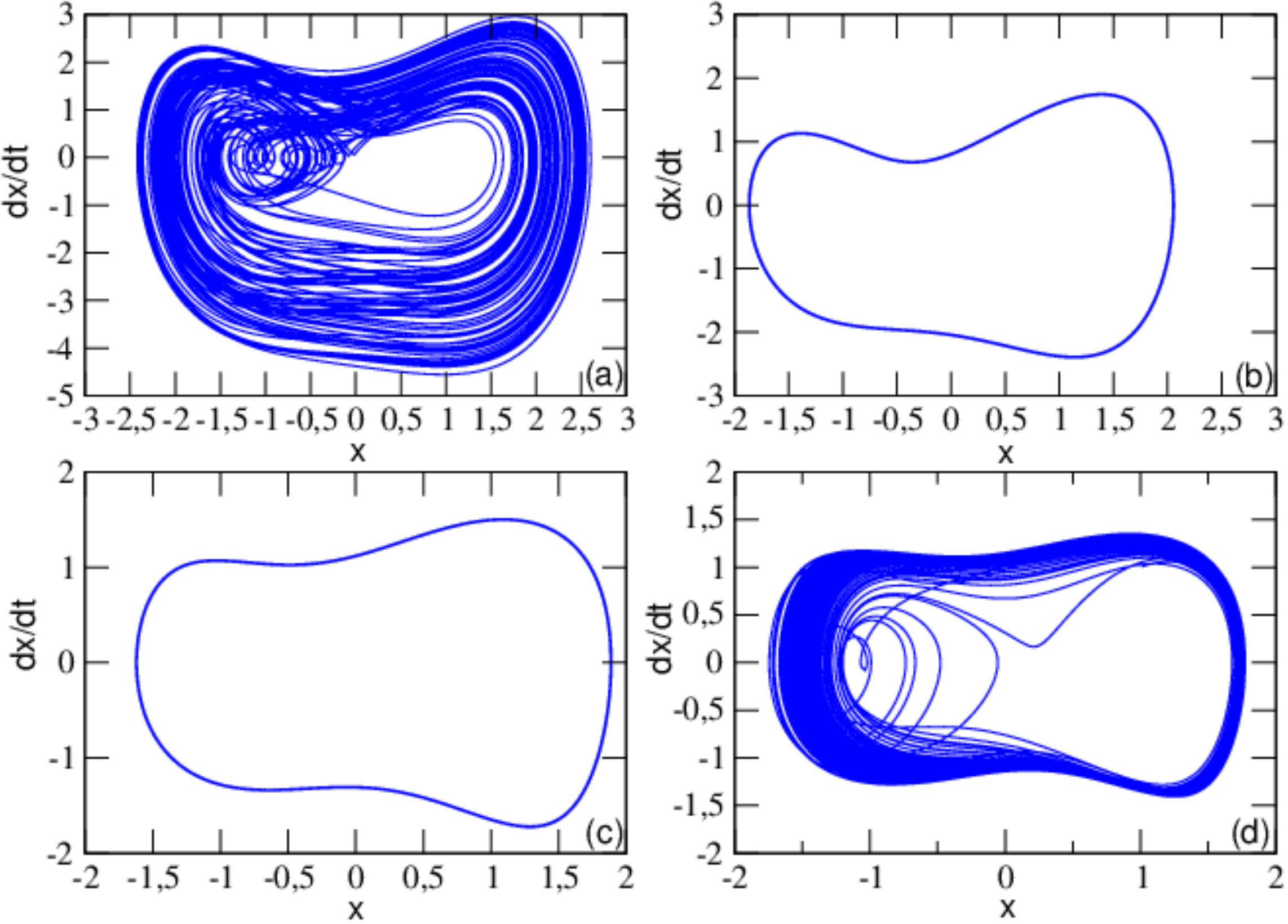}
\end{center}
\caption{ Phase portraits corresponding to Modified Rayleigh-Duffing oscillator with parameters of Fig.\ref{fig:5}; 
$(a) $ chaotic orbit $\mu=0.0001$, $(b)$  period-1 orbit $\mu=0.1$,
 $(c) $ period-1 orbit $\mu=0.5$, $(d)$ chaotic orbit $\mu=0.8$.}
\label{fig:7}
\end{figure}

\begin{figure}[htbp]
\begin{center}
 \includegraphics[width=12cm,  height=6cm]{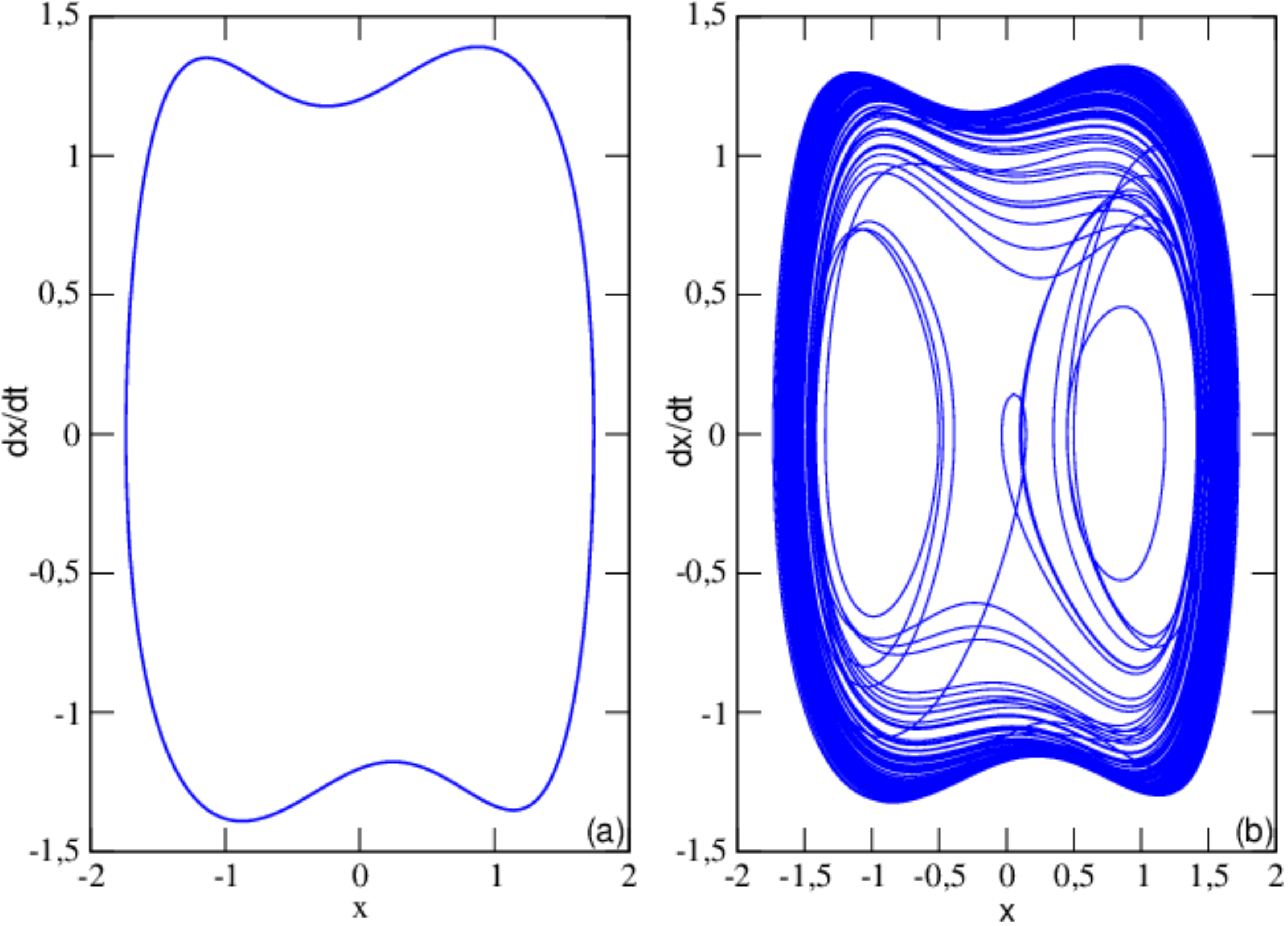}
\end{center}
\caption{ Phase portraits corresponding to Modified Rayleigh-Duffing oscillator when the modified damping parameters equals 
to $0$ with parameters of Fig.\ref{fig:6}; $(a) $ period-1 orbit $\mu=0.5$, $(b)$ chaotic orbit $\mu=0.695$.}
\label{fig:8}
\end{figure}

In this part, we study the behavior of the system given by Eq.\ref{eq.1} as a function of the damping parameter for different values
of the external perturbation. The bifurcation diagram and the maximal Lyapunov exponents have been represented for
the variable $x$, and they can be seen in Fig.\ref{fig:5}. A positive Lyapunov exponent for a bounded attractor is usually a sign of
chaos. We want to check the threshold of the external amplitude for the onset of possible chaos obtained in Section 3.
For $\Omega = 1$, the critical value of the external force has been obtained numerically for $ F_{cr}= 0.5$. Above this value, 
numerical simulations have been carried out for the selected parameter values $F = 0.5$ (see Figs.\ref{fig:5} $(a)$ and $(c)$) and $F = 0.6$ (see
Figs. \ref{fig:5}$(b)$ and $(d)$). From these figures, one can see that the thresholds of damping amplitude for the onset of chaos increase
when the external amplitude increases above $ F_{cr}$. After the chaotic motion in the small domain of $\mu$ ($[0,0.085] $ for $F=0.5$ and $[0,0.13]$
 for $F=0.6$),
 the Lyapunov exponent changes from a negative value to a positive value when $\mu$ increases, signifying the appearance of homoclinic chaos motion. From
Figs. \ref{fig:3} and \ref{fig:5}, we 
can noticed that the Melnikov critrical value $\mu_{cr}$ obtained in Section 3 is confirmed by numerical simulations. 
 The phase portrait of chaotic and periodic
orbits have been plotted in Fig. \ref{fig:7} with parameters of Fig.\ref{fig:5}. Clearly, we noticed that periodic appear when 
$\mu=0.085$ and perxist in means forms and is destroyed when this parameter is increasing which indicate
the homoclinic chaos is appeared.  Fig.\ref{fig:6} illustrate the bifurcation diagram and the maximal
 Lyapunov exponents
of our system when the modified parameters equals $0$ $(\alpha=0, \beta=0, k_1=0, k_2=0)$  and its corresponding phase portrait have been plotted in 
Fig.\ref{fig:8}. These figures show that our results coincide exactly with the results when the modified parameters equals $0$
 which are obtained for Rayleigh-Duffing by Siewe Siewe and al.(see \cite{2}).  The effect of nonlinear damping parameters, parametric excitation and 
external forced amplitude are also seeked through these figures. 

\begin{figure}[htbp]
\begin{center}
 \includegraphics[width=12cm,  height=8cm]{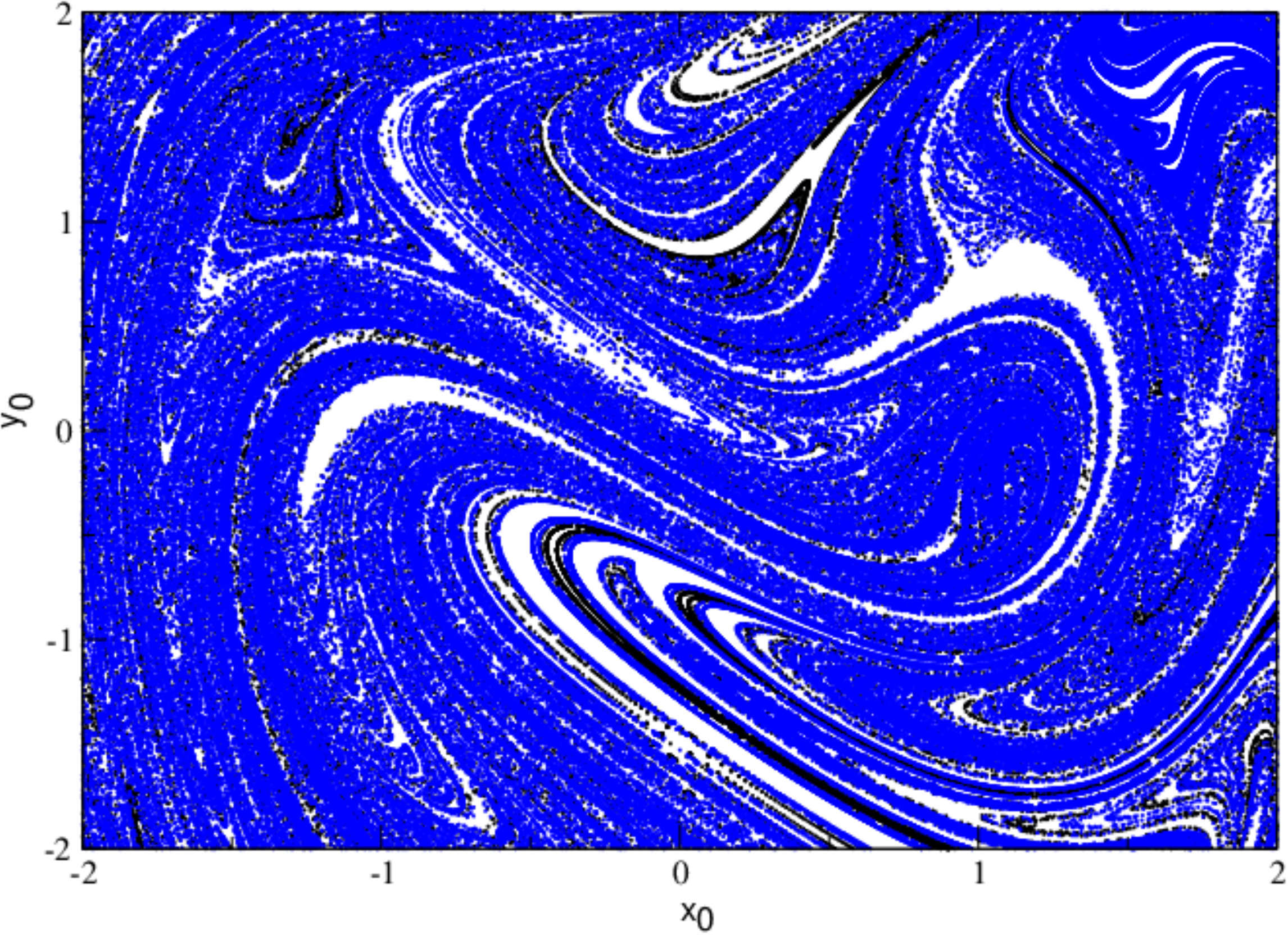}
\end{center}
\caption{Basin of attraction corresponding to the system with $\mu=0.0001$, the
others parameters are: $\gamma=-1, \lambda=1,\alpha=0.3, \beta=0.05, \Omega=1, \epsilon=-1, k_1=0.5, k_2=0.05$ and $F=0.5$. }
\label{fig:9}
\end{figure}

\begin{figure}[htbp]
\begin{center}
 \includegraphics[width=12cm,  height=8cm]{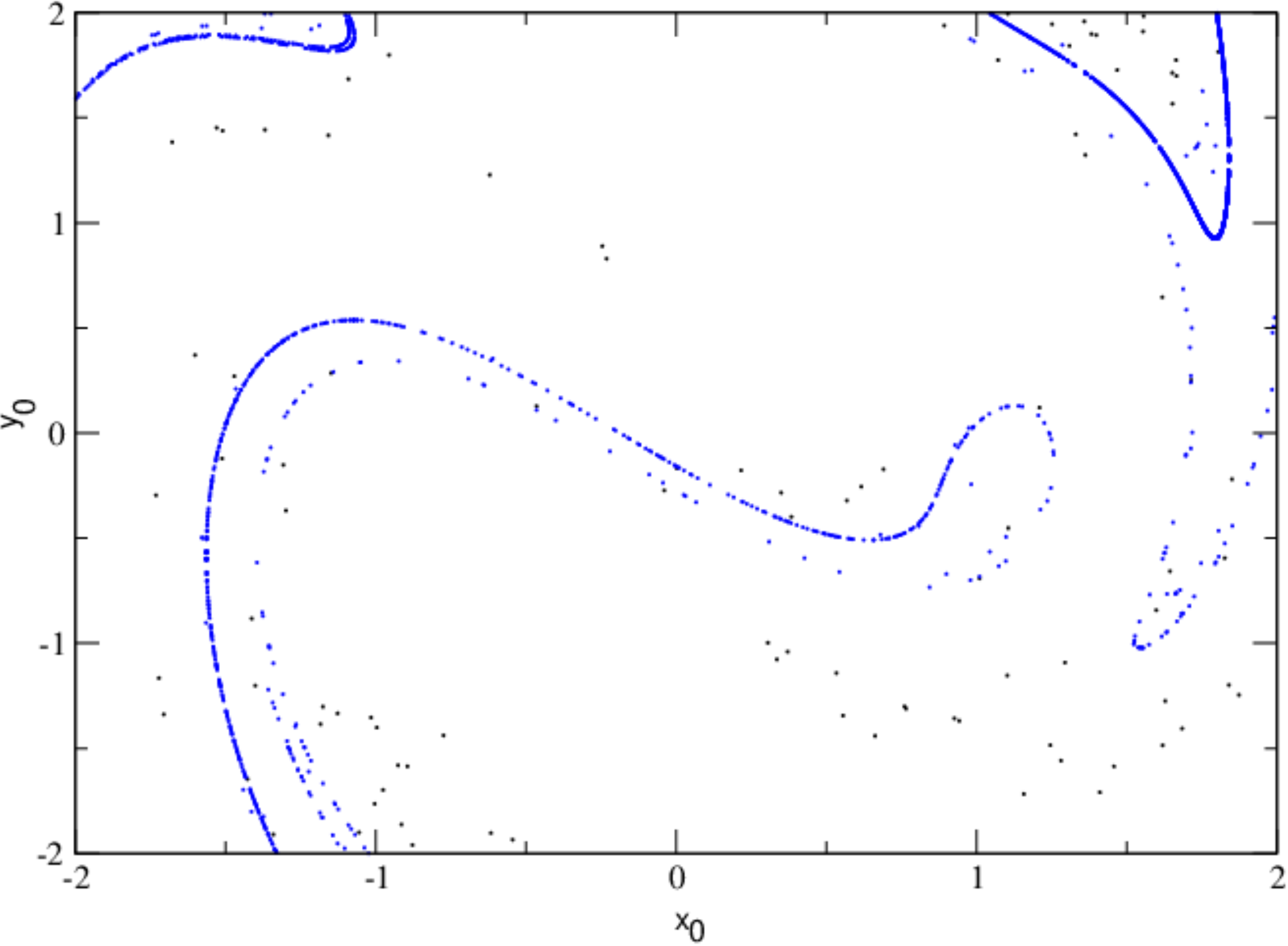}
\end{center}
\caption{Basin of attraction corresponding to the system with $\mu=0.045$, and the parameters of \ref{fig:9}.}
\label{fig:14}
\end{figure}

\begin{figure}[htbp]
\begin{center}
 \includegraphics[width=12cm,  height=8cm]{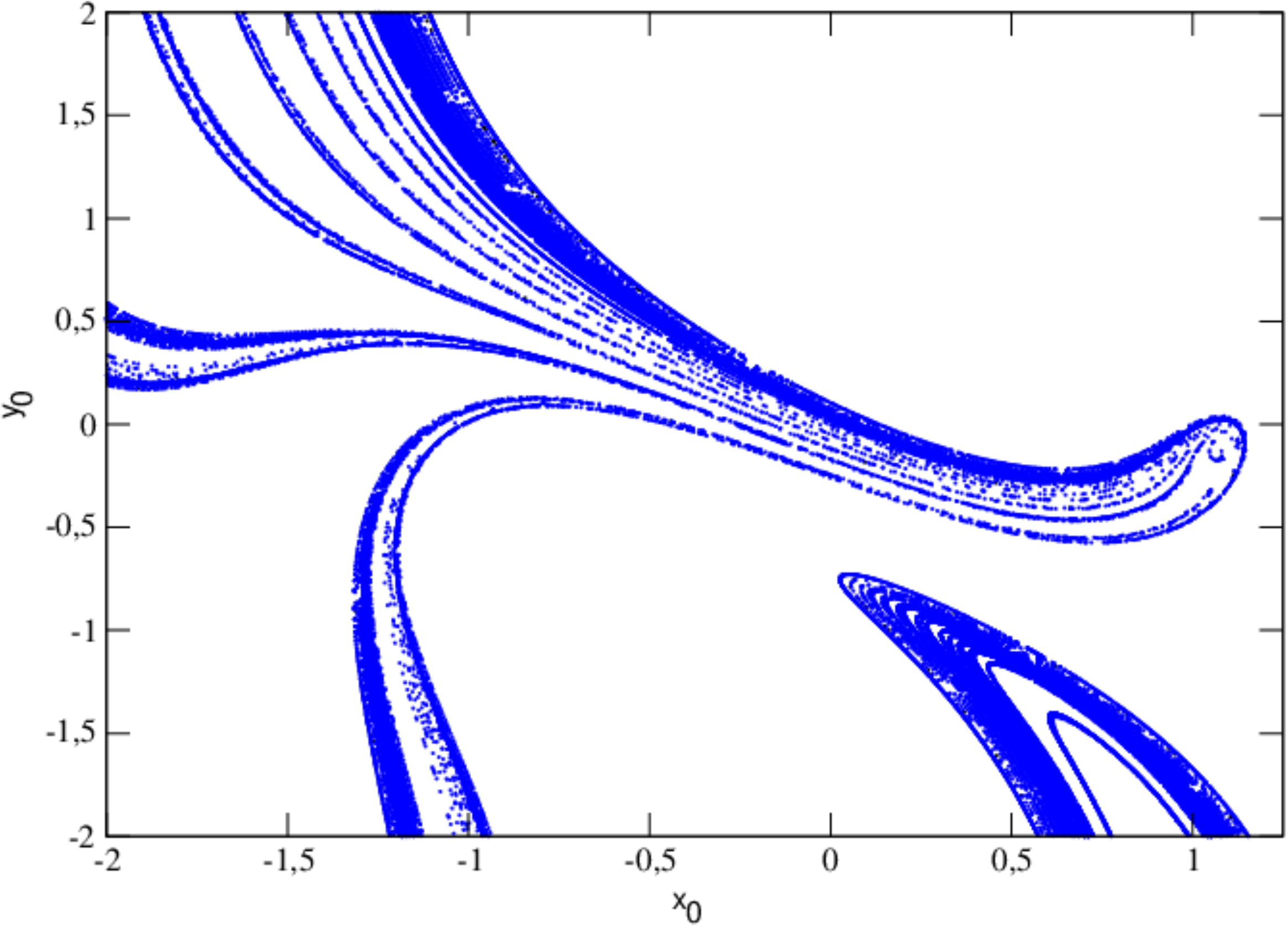}
\end{center}
\caption{Basin of attraction corresponding to the system with $\mu=0.6$, and the parameters of \ref{fig:9}.}
\label{fig:10}
\end{figure}

\begin{figure}[htbp]
\begin{center}
 \includegraphics[width=12cm,  height=8cm]{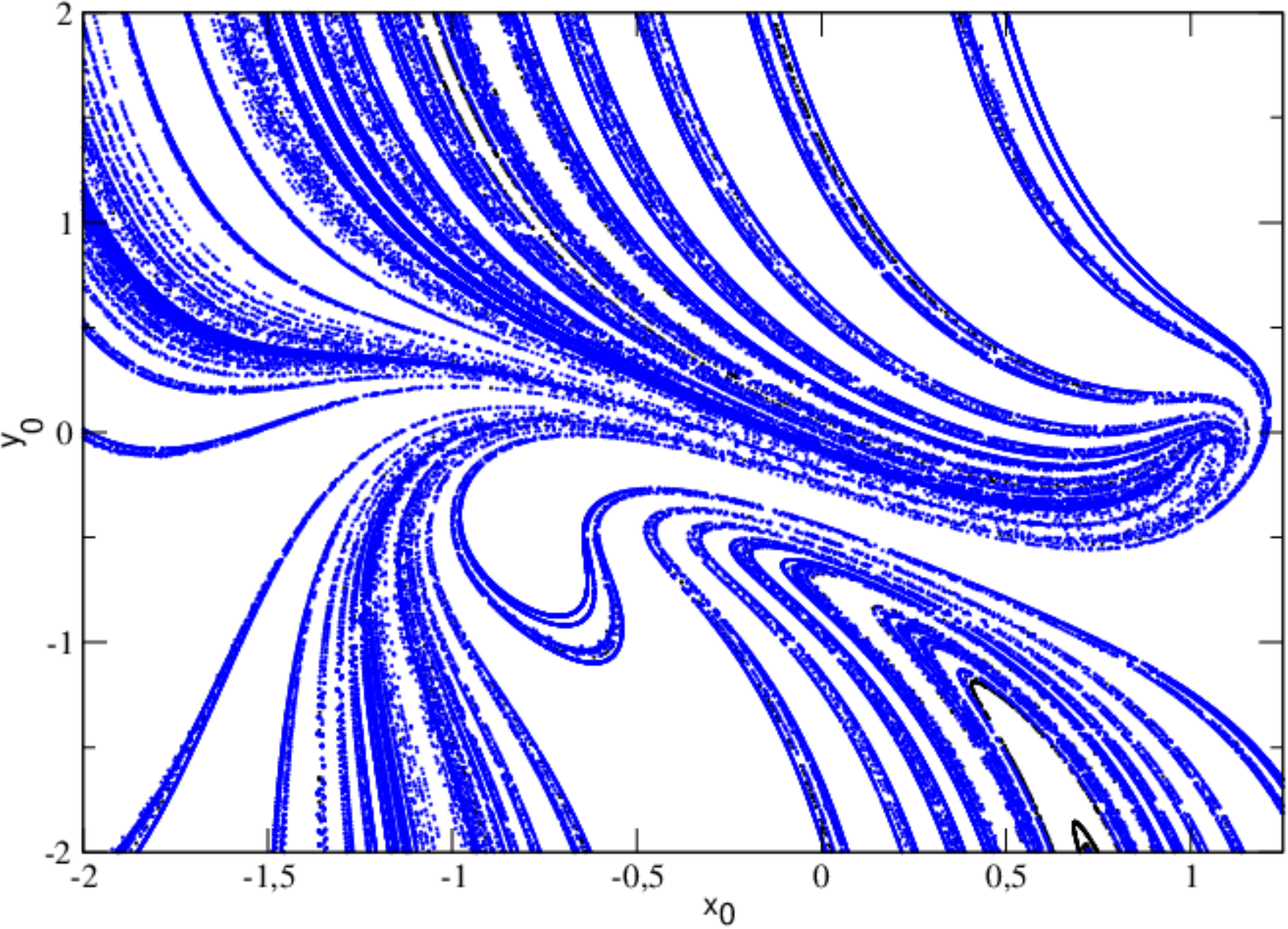}
\end{center}
\caption{Basin of attraction corresponding to the system with $\mu=0.8$, and the parameters of \ref{fig:9}.}
\label{fig:11}
\end{figure}

\begin{figure}[htbp]
\begin{center}
 \includegraphics[width=12cm,  height=8cm]{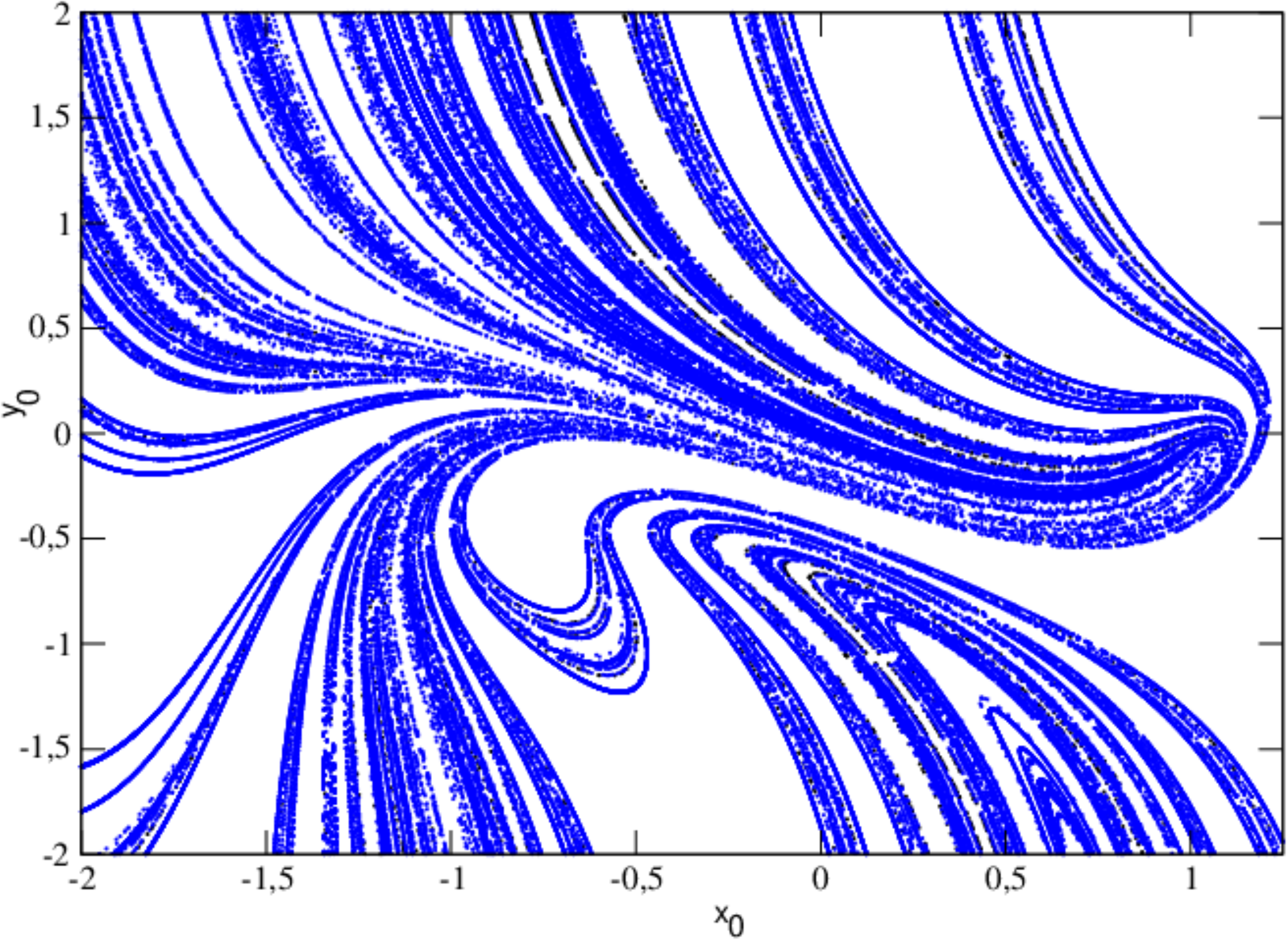}
\end{center}
\caption{Basin of attraction corresponding to the system with $\mu=0.85$, and the parameters of \ref{fig:9}.}
\label{fig:12}
\end{figure}

\begin{figure}[htbp]
\begin{center}
 \includegraphics[width=12cm,  height=8cm]{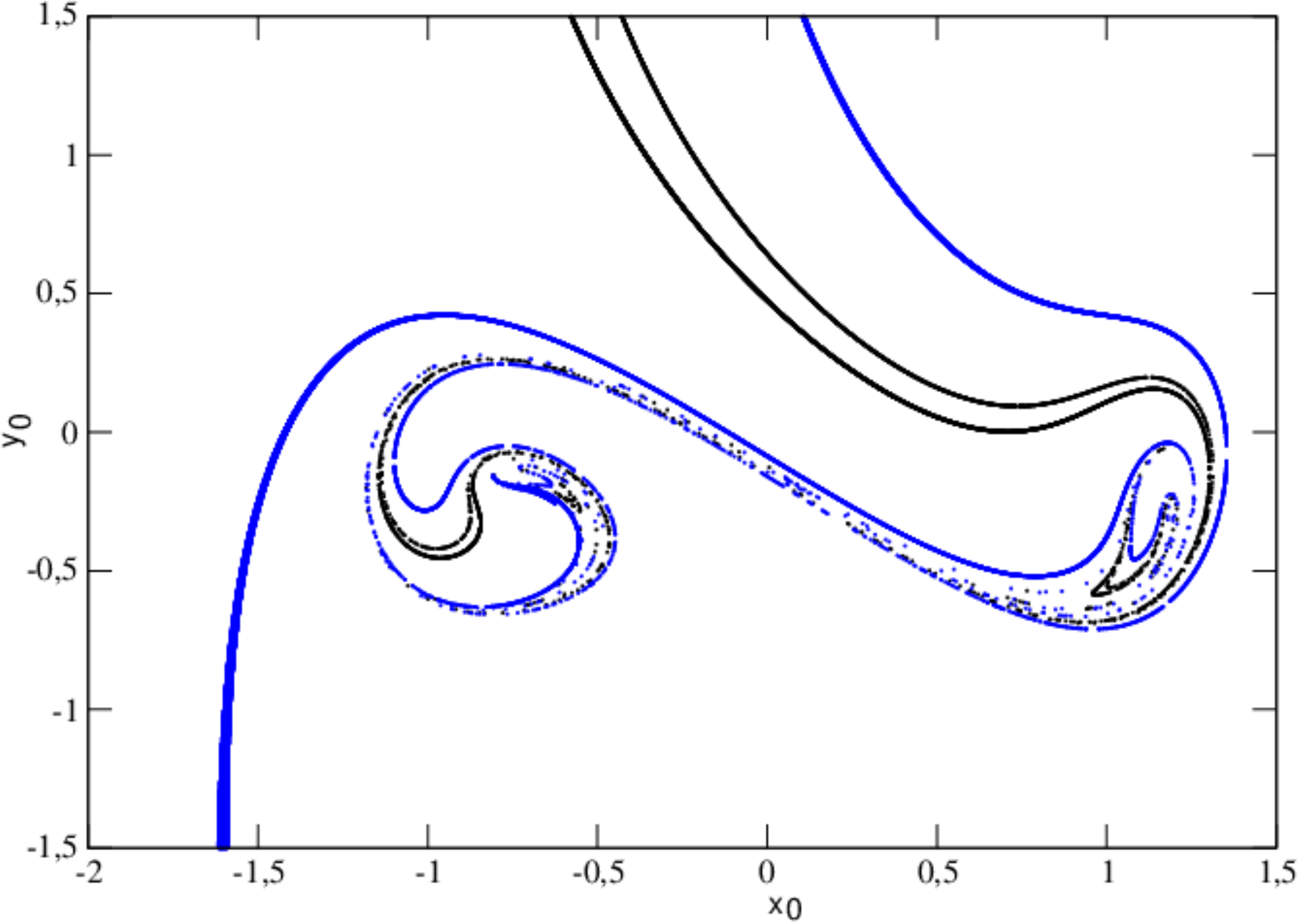}
\end{center}
\caption{Basin of attraction corresponding to the system with $\mu=0.6$,and the parameters of \ref{fig:8}.}
\label{fig:13}
\end{figure}

A basin of attraction is defined as the set of points taken as initial conditions, that are attracted to a
fixed point or an invariant set. The basin of attraction in this case signals the points in phase space
that are attracted to a safe oscillation within the potential well, and the set of points that escape
outside the potential well to the infinity.
 In order to verify the analytical results obtained in the previous sections, we have numerically integrated the system
by using a fourth order Runge–Kutta in order to investigate the homoclinic chaos in our model. We want to study what
is the effect of using the nonlinear damping terms on the equation of the oscillator and how the basins of attraction are
affected as the coefficient parameter $\mu$ is varied. To show the fractal structure, we consider the case of the bifurcation
close to the resonance since it may undergo the limit cycles in the system. We  see through from
Figs.\ref{fig:9}, \ref{fig:14}, \ref{fig:10}, \ref{fig:11}, \ref{fig:12}, \ref{fig:13} , the basin boundaries become fractal, which means that
 the damping parameter value $\mu$ has contributed to
the fractalization of the boundaries, with the corresponding uncertainty associated to this fact. For instance, as this control 
parameter increases above this critical value, the regular shape of
basin of attraction is destroyed and the fractal behavior becomes more and more visible (see Fig.\ref{fig:11} and \ref{fig:12}). Such fractal
boundaries indicate that whether the system is attracted to one or the other periodic attractor may be very sensitive
to initial conditions. It is also found that even if $\mu$ is increased beyond the analytical critical value for the homoclinic
bifurcation, it is still possible that the final steady motion could be periodic rather than chaotic. These results confirme our analytical result.
Finally, we prove through from Figs.  \ref{fig:12}  and \ref{fig:13} the modified parameters of habituelly Rayleigh-Duffing oscillator are very effect on chaotic motions 
of this system.

\newpage
\section{Conclusion}
In this paper, we have studied conditions of a global homoclinic bifurcation in
a double well potential modified Rayleigh-Duffing system with pure cubic nonlinear damping
coefficient term. Using the Melnikov method we have got the analytical formula for transition
to chaos in a one degree of freedom, system subjected to self-excitation term 
with a non-symmetric stiffness with parametric excitation. In our case this effect
is mutually introduced through the modified Rayleigh-Duffing damping and parametric excitation terms. The transition boundaries 
in the parameter space are obtained, which divide the space into different regions. In each region, the solutions are explored 
theoretically and numerically. The
critical value of damping coefficient $\mu$ under which the system oscillates chaotically has been estimated, in the first step
by means of the Melnikov method and later confirmed by calculating the corresponding Lyapunov exponent, bifurcation diagrams, 
and basin of attractions. Results were given for external periodic perturbation. By means of the basin of
attraction, we have shown that for certain regions of parameter space, the deterministic system driven harmonically
experiences behaviors that may be chaotic or non-chaotic.
The Melnikov method, is sensitive to a global homoclinic bifurcation and gives a necessary condition when the
damping coefficient $\mu = \mu_{cr1}$ is larger than the critical homoclinic bifurcation values. 
 Our analytical results are consistent with direct computations on homoclinic orbits. It is also investigated the effect of unpure
 quadratic nonlinear damping
and parametric excitation amplitude on chaotic behavior through the melnikov criteria and attraction basin.

\section*{Acknowlegments}
The authors thank IMSP-UAC and Benin gorvernment for financial support.

\section*{Appendix A} 
 Evaluation of the integrals from $I_0$, to $I_6$  is straightforward. After substitution $x_h (t)$, and
$y_h(t)$ (Eq. (\ref{eq.9})) we have:

 \begin{eqnarray}
  I_0&=&\frac{2\gamma^2}{\lambda}\int_{-\infty}^{+\infty}sech^2{(\sqrt{-\gamma}t)}\tanh^2{(\sqrt{-\gamma}t)}dt \nonumber\\
 &=&\frac{2\gamma^2}{\lambda}\int_{-\infty}^{+\infty}\frac{\tanh^2{(\sqrt{-\gamma}t)}}{\cosh^2{(\sqrt{-\gamma}t)}} dt.\label{eq.A1}
 \end{eqnarray}
and simple algebraic manipulations:
\begin{eqnarray}
 \tau=t\sqrt{-\gamma}, \quad \tanh\tau=\xi, \label{eq.A2}
\end{eqnarray}
we obtain 
\begin{eqnarray}
 I_0=\frac{2\gamma^2}{\lambda\sqrt{-\gamma}}\int_{-1}^{1}\xi^2d\xi.\label{eq.A3}
\end{eqnarray}
Finally the result of one has a following expression:
\begin{eqnarray}
  I_0=\frac{-4\gamma\sqrt{-\gamma}}{3\lambda}. \label{eq.A4}
\end{eqnarray}

\begin{figure}[htbp]
\begin{center}
 \includegraphics[width=8cm,  height=3cm]{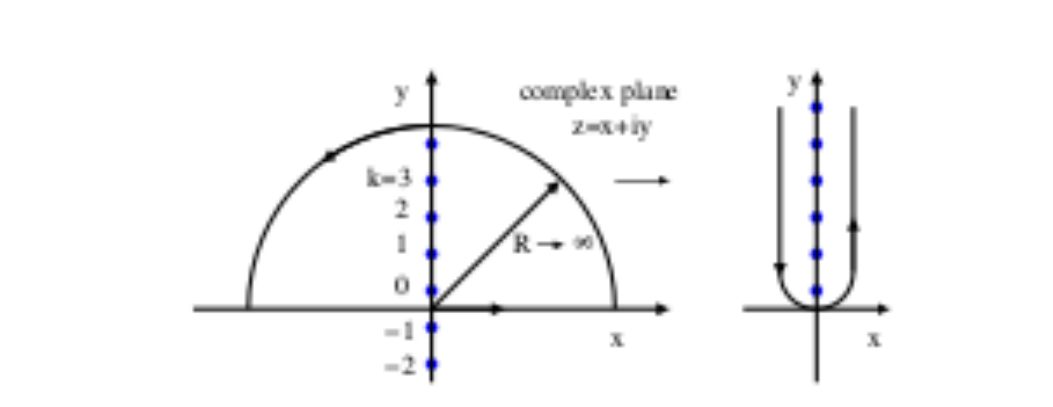}
\end{center}
\caption{Deformed contour integration schema and imaginary poles.}
\label{fig:A.1}
\end{figure}

After same algebraic manipulations, $I_1$, and $I_4$ become finally
\begin{eqnarray}
 I_1=-\frac{16\gamma^3\sqrt{-\gamma}}{35\lambda^2} \label{eq.A5}
\end{eqnarray}
and
\begin{eqnarray}
 I_4=0.  \label{eq.A6}
\end{eqnarray}
On the the hand the integrals $I_2$ and $I_3$ can be evaluated by using  following expressions:
\begin{eqnarray}
&& \int\frac{\sinh^p\tau}{\cosh^{2n+1}\tau}d\tau=\frac{\sinh^{p+1}\tau}{2n}\times\cr
&&[sech^{2n}\tau+
\sum_{k=1}^{n-1}\frac{(2n-p-1)(2n-p-3)...(2n-p-2k+1)}{2^k(n-1)(n-2)...(n-k)}sech^{2n-2k}]+\cr
&&\frac{(2n-p-1)(2n-p-3)...(3-p)(1-p)}{2^nn!}\int\frac{\sinh^p\tau}{\cosh\tau}d\tau \label{eq.A7}
\end{eqnarray}

\begin{eqnarray}
&& \int\frac{\sinh^p\tau}{\cosh^{2n}\tau}d\tau=\frac{\sinh^{p+1}\tau}{2n}\times\cr
&&[sech^{2n-1}\tau+
\sum_{k=1}^{n-1}\frac{(2n-p-2)(2n-p-4)...(2n-p-2k)}{(2n-3)(2n-5)...(2n-2k-1)}sech^{2n-2k-1}]+\cr
&&\frac{(2n-p-2)(2n-p-4)...(-p+2)(-p)}{(2n-1)!!}\int\sinh^p\tau d\tau \label{eq.A8}
\end{eqnarray}
\begin{eqnarray}
 \int\frac{\sinh^2\tau}{\cosh\tau}d\tau =\sinh\tau-\arctan(\sinh\tau), \label{eq.A9}
\end{eqnarray}
and
\begin{eqnarray}
 \int\sinh^3\tau d\tau=\frac{1}{3}\cosh^3\tau-\cosh\tau \label{eq.A10}
\end{eqnarray}
With these expressions, the finite values of $I_2$ and $I_3$ are:
\begin{eqnarray}
 I_3=\pm\frac{\pi\gamma^2}{4\lambda}\sqrt{\frac{2}{\lambda}}\label{eq.A11}
\end{eqnarray}
and
\begin{eqnarray}
 I_4=0.  \label{eq.A12}
\end{eqnarray}
Now, we evaluate $I_5$ and $I_6$. 
\begin{eqnarray}
I_5&=&\frac{2\gamma}{\lambda}\sqrt{-\gamma}\int_{-\infty}^{+\infty}sech^2{(\sqrt{-\gamma}t)}\tanh{(\sqrt{-\gamma}t)}\cos\Omega(t+t_0)dt, \label{eq.A13}
\end{eqnarray}
 \begin{eqnarray}
  I_5=\frac{2\gamma\sqrt{-\gamma}}{\lambda}\sin{\Omega t_0}\int_{-\infty}^{+\infty}\frac{\tanh{\tau}}{\cosh^2{\tau}}\sin{\frac{\Omega\tau}{\sqrt{-\gamma}}}dt \label{eq.A14}
 \end{eqnarray}
We put 
  \begin{eqnarray}
  I_5^A=\int_{-\infty}^{+\infty}\frac{\tanh{\tau}}{\cosh^2{\tau}}\sin{\frac{\Omega\tau}{\sqrt{-\gamma}}}dt \label{eq.A15}
 \end{eqnarray}
This integrale can be calculated by using the residue theorem
\begin{eqnarray}
 \oint f(z)dz=2\pi i\sum_{k=1}^{N}Res[f(z),z_k], \label{eq.A16}
\end{eqnarray}
where 
\begin{eqnarray}
 Res[f(z),z_k]=\frac{1}{(m-1)!}\lim_{z\rightarrow z_k}\frac{d^{m-1}}{z^{m-1}}[(z-z_k)^mf(z)]. \label{eq.A17}
\end{eqnarray}
In our case,
\begin{eqnarray}
 f(z)=4\frac{\exp(z)-\exp(-z)}{(\exp(z)+\exp(-z))^3}\exp\frac{i\Omega z}{\sqrt{-\gamma}}, \label{eq.A.18}
\end{eqnarray}
where on the real axis (Fig. \ref{fig:A.1})$Rez=\tau:$

\begin{eqnarray}
 Imf(z)=\frac{\tanh\tau}{\cosh^2\tau}\sin(\frac{\Omega\tau}{\sqrt{-\gamma}}).\label{eq.A19} 
\end{eqnarray}
The multiplicity of each pole of the complex function $f(z)$ (Eq. (\ref{eq.A.18}):
\begin{eqnarray}
 z_k=(\frac{\pi}{2}+\pi k)i \quad for k=1,2,... \label{eq.A20}  
\end{eqnarray}
can be easily determined as $ m=3$. After summation of all poles (Fig. \ref{fig:A.1}) we get:
 
 \begin{eqnarray}
  I_5^A=\frac{2\pi\Omega^2}{\gamma}\frac{\sin(\frac{\Omega z_0}{\sqrt{-\gamma}})}{\sinh(\frac{\Omega\pi}{2\sqrt{-\gamma}})} \label{eq.A21}
 \end{eqnarray}
Finally, the result of the above analysis can be written:
  
\begin{eqnarray}
  I_5=\frac{4\pi\Omega^2}{\lambda}\frac{\sin(\Omega t_0)}{\sinh(\frac{\Omega\pi}{2\sqrt{-\gamma}})} \label{eq.A22}
 \end{eqnarray}
Integral $ I_6$ is calculated with the same algebraic manipulations and the can be written as follows:
\begin{eqnarray}
 I_6=\mp\frac{23}{70}\pi\Omega\sqrt{\frac{2}{\lambda}}sech(\frac{\pi\Omega}{2\sqrt{-\gamma}})\sin\Omega t_0. \label{eq.A23} 
\end{eqnarray}

\end{document}